# Quantum Radiation Properties of Dirac Particles in General Nonstationary Black Holes

Jia-Chen Hua[1] Yong-Chang Huang [2, 3]

1. Department of Physics, University of Houston, Houston, Texas 77004, USA
2. Institute of Theoretical Physics, Beijing University of Technology, Beijing 100124, P. R. China
3. CCAST ( World Lab. ), P. O. Box 8730, Beijing, 100080, P. R. China

## Abstract

Quantum radiation properties of Dirac particles in general nonstationary black holes in the general case is investigated by both using the method of generalized tortoise coordinate transformation and considering the asymptotic behaviors of both the first and second order forms of Dirac equations near the event horizon. It is generally shown that the temperature and shape of event horizon of this kind of black holes depend on both the time and different angles. Further, we give a general expression of the new extra coupling effect in thermal radiation spectrum of Dirac particles which is missing in that of scalar particles. Also, we reveal a relationship that is ignored before between thermal radiation and non-thermal radiation in the case of scalar particles, which is that the chemical potential in thermal radiation spectrum is equal to the highest energy of the negative energy state of scalar particles in non-thermal radiation for general nonstationary black holes.

**Key words:** Dirac equation, nonstationary black holes, radiation, tortoise coordinate transformation, spin-rotation coupling effect, spin-acceleration coupling effect

**PACS numbers: 04.70.Dy, 97.60.Lf**

## 1. Introductions

Hawking's original discovery of quantum thermal radiation of black holes has been extensively studied in the fourth quarter of the last century [1]. An important subject in black hole physics is to reveal the thermal and non-thermal properties of various black holes. The last few decades has witnessed much progress in investigating the thermal radiation and non-thermal radiation of scalar fields or Dirac particles in many different kinds of black holes. Nevertheless, most of these researches had concentrated on static or stationary concrete black holes (e.g., [2-4]) since the original derivation of black hole evaporation involves gravitational collapse and is technically rather complicated [1]. It is, however, possible to understand the particle emission within the approaches that do not depend heavily on the gravitational collapse itself. Thus in 1976, Damour and Ruffini presented a treatment in which the gravitational field is independent of time [5]. In their approach the particle emission arises directly from a quantum mechanical barrier penetration across the event horizon of black holes. Sannan improved their method and obtained the probability distributions of both bosons and fermions emitted to infinity [6]. Then in the 1990s, Refs. [7-9] made further improvements in these parts, and calculated the location of the event horizon and the temperature of nonstationary black holes simultaneously by using the method of generalized tortoise coordinate transformation, and obtained the Hawking thermal spectrum. Ref. [7-9]'s results are consistent with those obtained by calculating the vacuum expectation values of the renormalized energy-momentum tensors [10, 11] on some spherically symmetric nonstationary black holes. Therefore, one can obtain the event horizon equation, Hawking temperature and thermal radiation spectrum of nonstationary black holes more conveniently and more exactly via the method of generalized tortoise coordinate transformation [12-14].

However, it is very difficult to deal with the quantum thermal effect of Dirac particles in nonstationary black holes. The difficulty mainly lies in the non-separability of variables for Chandrasekhar-Dirac equations [15] in the nonstationary axisymmetric and more general spacetime. Recently, Refs. [16-18] have suggested that considering simultaneously the asymptotic behaviors of the first order and second order forms of Dirac equation near the event horizon would overcome this difficult problem. In Ref. [16-18], under the generalized tortoise coordinate transformation, each second order equation induced from Chandrasekhar-Dirac equation takes the standard form of wave equation near the event horizon, to which separation of variables is possible. The location and the temperature of the event horizon are just the same as those obtained in the case of the thermal radiation of scalar particles in some nonstationary spacetime. Moreover, a kind of new term, representing a new extra spin-rotation coupling effect in Ref. [16, 17] and a new extra spin-acceleration coupling effect in Ref. [18], appears in the Fermionic spectrum of Dirac particles. This kind of new term is absent from the Bosonic spectrum of scalar particles. However, these researches were restricted to concrete black holes and did not notice the relation between thermal radiation and non-thermal radiation in the case of scalar particles.

Refs. [19-21] investigated thermodynamics of black holes in lovelock gravity and in AdS/dS spaces. One of the key issues of these researches is the calculation of Hawking temperature. By using the method of generalized tortoise coordinate transformation, this letter may provide an alternative and convenient way to obtain the Hawking temperature in these researches.

The main purposes of this letter are to investigate the thermal and non-thermal radiation of Dirac particles in general nonstationary black holes and the relation between the two kinds of radiation in the case of scalar particles, and give a general expression of the new extra coupling effect in thermal radiation spectrum of Dirac particles. Section 2 calculates the location of event horizon, Hawking temperature and the thermal radiation spectrum of Dirac particles in detail by using the method of generalized tortoise coordinate transformation [7-9, 12-14] and considering simultaneously the asymptotic behaviors of the first order and second order forms of Dirac equation near the event horizon [16-18]. In section 3, we formulate the highest energy of negative energy state of Dirac particles in non-thermal radiation by using the methods and the conclusions of references [22-28]. Section 4 is the discussion. In this section, we analyse the thermal radiation spectrum of Dirac particles in general nonstationary black holes, give the general expression of the new extra coupling effect, and discuss the relationship between thermal radiation and non-thermal radiation of black holes in the case of scalar particles. The last section is summary and conclusion.

## 2. Calculation of the thermal radiation in detail

The square of infinitesimal line element of the most general spacetime can be written as

$$ds^2 = g_{\mu\nu} dx^\mu dx^\nu \,, \tag{1}$$

where we take the advanced Eddington-Finkelstein coordinates ($x^0 = v\,,\ x^1 = r\,,\ x^2 = \theta\,,\ x^3 = \varphi$), and make the conventions that all indices of Latin letters $j,k = 0,2,3$ and all indices of Greek letters $\mu,\nu = 0,1,2,3$.

The surface equation of event horizon can be written as $F(v,r,\theta,\varphi)=0$, where $r_H = r_H(v,\theta,\varphi)$, which should satisfy null surface condition

$$g^{\mu\nu}\frac{\partial F}{\partial x^{\mu}}\frac{\partial F}{\partial x^{\nu}}=0\,. \tag{2}$$

From the above equation, we can obtain

$$g^{11}-2g^{1j}r_{,j}+g^{jk}r_{,j}r_{,k}=0\,, \tag{3}$$

$r_H$ is the location of event horizon and depends on the time and different angles.

To write out the explicit form of Dirac equation in the Newman-Penrose (NP) formalism [29], we establish the following complex null-tetrad system $\{l^{\mu},n^{\mu},m^{\mu},\bar{m}^{\mu}\}$ at each point in a 4-dimensional spacetime, where $l^{\mu}$ and $n^{\mu}$ are a pair of standard real null vectors such that $n^{\mu}l_{\mu}=1$, $m^{\mu}$ is the complex null vector, and the fourth vector $\bar{m}^{\mu}$ is just the complex conjugate of $m^{\mu}$. The NP tetrad satisfies the orthogonal conditions

$$\begin{cases} l_{\mu}l^{\mu}=n_{\mu}n^{\mu}=m_{\mu}m^{\mu}=\bar{m}_{\mu}\bar{m}^{\mu}=0 \\ l_{\mu}n^{\mu}=-m_{\mu}\bar{m}^{\mu}=1 \\ l_{\mu}m^{\mu}=l_{\mu}\bar{m}^{\mu}=n_{\mu}m^{\mu}=n_{\mu}\bar{m}^{\mu}=0 \end{cases}, \tag{4}$$

and

$$\begin{cases} g^{\mu\nu}=l^{\mu}n^{\nu}+n^{\mu}l^{\nu}-m^{\mu}\bar{m}^{\nu}-\bar{m}^{\mu}m^{\nu} \\ g_{\mu\nu}=l_{\mu}n_{\nu}+n_{\mu}l_{\nu}-m_{\mu}\bar{m}_{\nu}-\bar{m}_{\mu}m_{\nu} \end{cases}, \tag{5}$$

and then we obtain the corresponding directional derivatives

$$\begin{cases} D=l^{\mu}\dfrac{\partial}{\partial x^{\mu}} \\ \Delta=n^{\mu}\dfrac{\partial}{\partial x^{\mu}} \\ \delta=m^{\mu}\dfrac{\partial}{\partial x^{\mu}} \\ \bar{\delta}=\bar{m}^{\mu}\dfrac{\partial}{\partial x^{\mu}} \end{cases}. \tag{6}$$

The dynamical behavior of spin-1/2 particles in curved spacetime is described by the four coupled Chandrasekhar-Dirac equations [15] expressed in the Newman-Penrose formalism

$$\left(D+\varepsilon-\rho+ieA_{\mu}l^{\mu}\right)F_1+\left(\bar{\delta}+\pi-\alpha+ieA_{\mu}\bar{m}^{\mu}\right)F_2-\frac{1}{\sqrt{2}}i\mu_0G_1=0 \tag{7}$$

$$\left(\Delta+\mu-\gamma+ieA_{\mu}n^{\mu}\right)F_2+\left(\delta+\beta-\tau+ieA_{\mu}m^{\mu}\right)F_1-\frac{1}{\sqrt{2}}i\mu_0G_2=0 \tag{8}$$

$$\left(D+\bar{\varepsilon}-\bar{\rho}+ieA_{\mu}l^{\mu}\right)G_2-\left(\delta+\bar{\pi}-\bar{\alpha}+ieA_{\mu}m^{\mu}\right)G_1-\frac{1}{\sqrt{2}}i\mu_0F_2=0 \tag{9}$$

$$\left(\Delta+\bar{\mu}-\bar{\gamma}+ieA_{\mu}n^{\mu}\right)G_1-\left(\bar{\delta}+\bar{\beta}-\bar{\tau}+ieA_{\mu}\bar{m}^{\mu}\right)G_2-\frac{1}{\sqrt{2}}i\mu_0F_1=0\,, \tag{10}$$

where $\mu_0$, $e$ are the mass and charge of the Dirac particles, respectively. $F_1,F_2,G_1,G_2$ are the four components of Dirac spinor in the Newman-Penrose formalism. $\varepsilon,\rho,\pi,\alpha,\mu,\gamma,\beta,\tau$ are spin coefficients introduced by Newman and Penrose [29], which satisfy

$$\begin{aligned}
\varepsilon &= \frac{1}{2}\left(l_{\mu;\nu}n^\mu l^\nu - m_{\mu;\nu}\overline{m}^\mu l^\nu\right)\\
\rho &= l_{\mu;\nu}m^\mu \overline{m}^\nu\\
\pi &= -n_{\mu;\nu}\overline{m}^\mu l^\nu\\
\alpha &= \frac{1}{2}\left(l_{\mu;\nu}n^\mu \overline{m}^\nu - m_{\mu;\nu}\overline{m}^\mu \overline{m}^\nu\right)\\
\mu &= -n_{\mu;\nu}\overline{m}^\mu m^\nu \qquad\qquad ,\\
\gamma &= \frac{1}{2}\left(l_{\mu;\nu}n^\mu n^\nu - m_{\mu;\nu}\overline{m}^\mu n^\nu\right)\\
\beta &= \frac{1}{2}\left(l_{\mu;\nu}n^\mu m^\nu - m_{\mu;\nu}\overline{m}^\mu m^\nu\right)\\
\tau &= l_{\mu;\nu}m^\mu n^\nu
\end{aligned}$$

and $\overline{\varepsilon},\overline{\rho},\overline{\pi},\overline{\alpha},\overline{\mu},\overline{\gamma},\overline{\beta},\overline{\tau}$ are complex conjugates of $\varepsilon,\rho,\pi,\alpha,\mu,\gamma,\beta,\tau$.

To investigate the thermal radiation of spin-1/2 particles, we to need deal with the behavior of the second order Dirac equations near the event horizon. It is consistent to consider the asymptotic behavior of both the first and second order Dirac equations at the same time since the four-component Dirac spinors should satisfy both of them. By substituting Eq. (7) and Eq. (8) into Eq. (9) and Eq. (10), one can obtain the second order form of Dirac equations for $(F_1, F_2)$ components as follows

$$\begin{aligned}
&-2\left(\Delta + \overline{\mu} - \overline{\gamma} + ieA_\mu n^\mu\right)\left[\left(D + \varepsilon - \rho + ieA_\mu l^\mu\right)F_1 + \left(\overline{\delta} + \pi - \alpha + ieA_\mu \overline{m}^\mu\right)F_2\right]\\
&+2\left(\overline{\delta} + \overline{\beta} - \overline{\tau} + ieA_\mu \overline{m}^\mu\right)\left[\left(\Delta + \mu - \gamma + ieA_\mu n^\mu\right)F_2 + \left(\delta + \beta - \tau + ieA_\mu m^\mu\right)F_1\right] - {\mu_0}^2 F_1 = 0\\
&-2\left(D + \overline{\varepsilon} - \overline{\rho} + ieA_\mu l^\mu\right)\left[\left(\Delta + \mu - \gamma + ieA_\mu n^\mu\right)F_2 + \left(\delta + \beta - \tau + ieA_\mu m^\mu\right)F_1\right]\\
&+2\left(\delta + \overline{\pi} - \overline{\alpha} + ieA_\mu m^\mu\right)\left[\left(D + \varepsilon - \rho + ieA_\mu l^\mu\right)F_1 + \left(\overline{\delta} + \pi - \alpha + ieA_\mu \overline{m}^\mu\right)F_2\right] - {\mu_0}^2 F_2 = 0
\end{aligned} \quad .(11)$$

Introducing the generalized tortoise coordinate transformation [5-9, 12-14, 16-18, 30]

$$\begin{cases}
r_* = r + \dfrac{1}{2\kappa}\ln\left[r - r_H(v,\theta,\varphi)\right]\\
v_* = v - v_0\\
\theta_* = \theta - \theta_0\\
\varphi_* = \varphi - \varphi_0
\end{cases}, \qquad (12)$$

and

$$\begin{cases}
\dfrac{\partial}{\partial r} = \dfrac{\partial}{\partial r_*} + \dfrac{1}{2\kappa(r - r_H)}\dfrac{\partial}{\partial r_*}\\
\dfrac{\partial}{\partial x^j} = \dfrac{\partial}{\partial {x_*}^j} - \dfrac{r_{H,j}}{2\kappa(r - r_H)}\dfrac{\partial}{\partial r_*}
\end{cases},$$

along with

$$\begin{cases}
\dfrac{\partial^2}{\partial r^2} = \dfrac{\left[2\kappa(r - r_H) + 1\right]^2}{\left[2\kappa(r - r_H)\right]^2}\dfrac{\partial^2}{\partial {r_*}^2} - \dfrac{1}{2\kappa(r - r_H)^2}\dfrac{\partial}{\partial r_*}\\
\dfrac{\partial^2}{\partial r\partial x^j} = \dfrac{2\kappa(r - r_H) + 1}{2\kappa(r - r_H)}\dfrac{\partial^2}{\partial r_*\partial {x_*}^j} + \dfrac{r_{H,j}2\kappa}{\left[2\kappa(r - r_H)\right]^2}\dfrac{\partial}{\partial r_*} - \dfrac{r_{H,j}\left[2\kappa(r - r_H) + 1\right]}{\left[2\kappa(r - r_H)\right]^2}\dfrac{\partial^2}{\partial {r_*}^2}\\
\dfrac{\partial^2}{\partial x^j\partial x^k} = \dfrac{\partial^2}{\partial {x_*}^j\partial {x_*}^k} - \dfrac{r_{H,j}}{2\kappa(r - r_H)}\dfrac{\partial^2}{\partial r_*\partial {x_*}^k} - \dfrac{r_{H,k}}{2\kappa(r - r_H)}\dfrac{\partial^2}{\partial r_*\partial {x_*}^j} + \dfrac{r_{H,j}r_{H,k}}{\left[2\kappa(r - r_H)\right]^2}\dfrac{\partial^2}{\partial {r_*}^2} - \dfrac{2\kappa(r - r_H)r_{H,jk} + 2\kappa r_{H,j}r_{H,k}}{\left[2\kappa(r - r_H)\right]^2}\dfrac{\partial}{\partial r_*}
\end{cases},$$

where $v_0,\theta_0,\varphi_0$ are parameters under the tortoise transformation, $\kappa \equiv \kappa(v_0,\theta_0,\varphi_0)$ is an adjustable parameter that depends on time and angular coordinates.

The methods of generalized tortoise coordinate transformation can simultaneously obtain the exact values of the location and the temperature of the event horizon of nonstationary black holes. Basically, this method is to reduce Klein-Gordon or Dirac equation in a known black hole background to a standard wave equation outside and near the event horizon by generalizing the common tortoise-type coordinate $r_* = r + \frac{1}{2\kappa}\ln(r - r_H)$ in a static or stationary spacetime [5, 6] (where $\kappa$ is the surface gravity of the studied event horizon) to a similar form in a nonstatic or nonstationary spacetime and by allowing the location of the event horizon $r_H$ to be a function of the advanced time $v = t + r_*$ and the angles $\theta, \varphi$. The tortoise coordinates only describe the spacetime outside the event horizon, and in this condition, $r_*$ tends to positive infinity when approaching to the infinite point, and $r_*$ tends to negative infinity at the event horizon. In fact, this felicitously describes the spacetime outside the event horizon that we want to investigate, in which the tortoise coordinates do not result in the singularities ( when we try to get the Klein-Gordon equation near the event horizon, we can take the limit of $r_*$ tending to negative infinity, i.e. $r \to r_H$, then we obtain the definitely finite form of the Klein-Gordon equation near the event horizon ). And also, for the spacetime inside the event horizon, we can extend the outgoing wave function by analytic continuation into the event horizon through the lower-half complex r-plane (See the paragraph before Eq. (23) for details).

Applying the generalized tortoise coordinate transformation to Eq. (7-10) and taking the limit of $r \to r_H$ (here and hereafter, $r \to r_H$ represents $v \to v_0$, $\theta \to \theta_0$, $\varphi \to \varphi_0$, $r \to r_H(v_0, \theta_0, \varphi_0)$), we can get

$$\begin{cases} \dfrac{\partial F_1}{\partial r_*} = \left.\dfrac{\bar{m}^1 - \bar{m}^j r_{H,j}}{l^j r_{H,j} - l^1}\right|_{r\to r_H} \dfrac{\partial F_2}{\partial r_*} \\ \dfrac{\partial F_2}{\partial r_*} = \left.\dfrac{m^1 - m^j r_{H,j}}{n^j r_{H,j} - n^1}\right|_{r\to r_H} \dfrac{\partial F_1}{\partial r_*} \\ \dfrac{\partial G_2}{\partial r_*} = \left.\dfrac{m^1 - m^j r_{H,j}}{l^1 - l^j r_{H,j}}\right|_{r\to r_H} \dfrac{\partial G_1}{\partial r_*} \\ \dfrac{\partial G_1}{\partial r_*} = \left.\dfrac{\bar{m}^1 - \bar{m}^j r_{H,j}}{n^1 - n^j r_{H,j}}\right|_{r\to r_H} \dfrac{\partial G_2}{\partial r_*} \end{cases}. \tag{13}$$

Then applying the generalized tortoise coordinate transformation to Eq. (11), via some arrangement, multiplying $\frac{2\kappa(r - r_H)}{g^{01}[2\kappa(r - r_H) + 1] - g^{0j} r_{H,j}}$ to both sides of the two second order equations for the coefficient of $\frac{\partial^2 F_1}{\partial r_* \partial v_*}$ and $\frac{\partial^2 F_2}{\partial r_* \partial v_*}$ to be 2 [7-9, 12-14, 16-18], finally taking the limit of $r \to r_H$ and substituting Eq. (13) into the two equations, we can obtain a united form of them

$$I\frac{\partial^2 \Psi}{\partial {r_*}^2} + 2\frac{\partial^2 \Psi}{\partial r_* \partial v_*} + B\frac{\partial^2 \Psi}{\partial r_* \partial \theta_*} + C\frac{\partial^2 \Psi}{\partial r_* \partial \varphi_*} + A_* \frac{\partial \Psi}{\partial r_*} = 0, \tag{14}$$

where

$$I = \lim_{r\to r_H} \frac{g^{11}[2\kappa(r - r_H) + 1]^2 - 2g^{1j} r_{H,j}[2\kappa(r - r_H) + 1] + g^{jk} r_{H,j} r_{H,k}}{2\kappa(r - r_H)[g^{01}[2\kappa(r - r_H) + 1] - g^{0j} r_{H,j}]},$$

$$B = 2\left.\frac{g^{12} - g^{j2} r_{H,j}}{g^{01} - g^{0j} r_{H,j}}\right|_{r\to r_H}, \quad C = 2\left.\frac{g^{13} - g^{j3} r_{H,j}}{g^{01} - g^{0j} r_{H,j}}\right|_{r\to r_H},$$

$$A_* = 2i(C_0 + C_1) + C_2 + C_{3*} + C_4 ,$$

$$C_0 = e\frac{A_\mu g^{\mu 1} - A_\mu g^{\mu j} r_{H,j}}{g^{01} - g^{0j} r_{H,j}}\bigg|_{r\to r_H} ,$$

$$C_{3*} = -2\kappa \frac{g^{11} - 2g^{1j} r_{H,j} + g^{jk} r_{H,j} r_{H,k}}{2\kappa(r - r_H)[g^{01}[2\kappa(r - r_H) + 1] - g^{0j} r_{H,j}]}\bigg|_{r\to r_H} ,$$

$$C_4 = \frac{-g^{jk} r_{H,jk}}{g^{01} - g^{0j} r_{H,j}}\bigg|_{r\to r_H} ,$$

$$\begin{aligned}
C_2 &= \frac{1}{g^{01} - g^{0j} r_{H,j}}\bigg|_{r\to r_H} \{\{2n^\mu l^1{}_{,\mu} - 2(\beta - \tau)\bar{m}^1 - 2(\bar{\beta} - \bar{\tau})m^1 + [(\varepsilon - \rho) + (\bar{\varepsilon} - \bar{\rho})]n^1 \\
&+ [(\bar{\mu} - \bar{\gamma}) + (\mu - \gamma)]l^1 - (\bar{m}^\mu m^1{}_{,\mu} + m^\mu \bar{m}^1_{,\mu})\} \\
&- \{2n^\mu l^j{}_{,\mu} - 2(\beta - \tau)\bar{m}^j - 2(\bar{\beta} - \bar{\tau})m^j + [(\varepsilon - \rho) + (\bar{\varepsilon} - \bar{\rho})]n^j \\
&+ [(\bar{\mu} - \bar{\gamma}) + (\mu - \gamma)]l^j - (\bar{m}^\mu m^j{}_{,\mu} + m^\mu \bar{m}^j_{,\mu})\} r_{H,j}\}\big|_{r\to r_H} \\
&+ \frac{1}{(g^{01} - g^{0j} r_{H,j})(n^j r_{H,j} - n^1)}\bigg|_{r\to r_H} \{\{[n^\mu \bar{m}^1_{,\mu} - \bar{m}^\mu n^1{}_{,\mu} + (\pi - \alpha)n^1 + (\bar{\mu} - \bar{\gamma})\bar{m}^1 - (\mu - \gamma)\bar{m}^1 - (\bar{\beta} - \bar{\tau})n^1] \\
&- [n^\mu \bar{m}^j_{,\mu} - \bar{m}^\mu n^j{}_{,\mu} + (\pi - \alpha)n^j + (\bar{\mu} - \bar{\gamma})\bar{m}^j - (\mu - \gamma)\bar{m}^j - (\bar{\beta} - \bar{\tau})n^j] r_{H,j}\}(m^1 - m^j r_{H,j}) \\
&+ \{[n^\mu m^1{}_{,\mu} - m^\mu n^1{}_{,\mu} + (\bar{\pi} - \bar{\alpha})n^1 + (\mu - \gamma)m^1 - (\bar{\mu} - \bar{\gamma})m^1 - (\beta - \tau)n^1] \\
&- [n^\mu m^j{}_{,\mu} - m^\mu n^j{}_{,\mu} + (\bar{\pi} - \bar{\alpha})n^j + (\mu - \gamma)m^j - (\bar{\mu} - \bar{\gamma})m^j - (\beta - \tau)n^j] r_{H,j}\}(\bar{m}^1 - \bar{m}^j r_{H,j})\}\big|_{r\to r_H}
\end{aligned} ,$$

$$\begin{aligned}
C_1 &= \frac{1}{g^{01} - g^{0j} r_{H,j}}\bigg|_{r\to r_H} \{\left\{\frac{[(\bar{\mu} - \bar{\gamma}) - (\mu - \gamma)]l^1}{2i} + \frac{[(\varepsilon - \rho) - (\bar{\varepsilon} - \bar{\rho})]n^1}{2i} - \frac{(\bar{m}^\mu m^1{}_{,\mu} - m^\mu \bar{m}^1_{,\mu})}{2i}\right\} \\
&- \left\{\frac{[(\bar{\mu} - \bar{\gamma}) - (\mu - \gamma)]l^j}{2i} + \frac{[(\varepsilon - \rho) - (\bar{\varepsilon} - \bar{\rho})]n^j}{2i} - \frac{(\bar{m}^\mu m^j{}_{,\mu} - m^\mu \bar{m}^j_{,\mu})}{2i}\right\} r_{H,j}\}\bigg|_{r\to r_H} \\
&+ \frac{1}{(g^{01} - g^{0j} r_{H,j})(n^j r_{H,j} - n^1)}\bigg|_{r\to r_H} \frac{1}{2i}\{\{[n^\mu \bar{m}^1_{,\mu} - \bar{m}^\mu n^1{}_{,\mu} + (\pi - \alpha)n^1 + (\bar{\mu} - \bar{\gamma})\bar{m}^1 - (\mu - \gamma)\bar{m}^1 - (\bar{\beta} - \bar{\tau})n^1] \\
&- [n^\mu \bar{m}^j_{,\mu} - \bar{m}^\mu n^j{}_{,\mu} + (\pi - \alpha)n^j + (\bar{\mu} - \bar{\gamma})\bar{m}^j - (\mu - \gamma)\bar{m}^j - (\bar{\beta} - \bar{\tau})n^j] r_{H,j}\}(m^1 - m^j r_{H,j}) \\
&- \{[n^\mu m^1{}_{,\mu} - m^\mu n^1{}_{,\mu} + (\bar{\pi} - \bar{\alpha})n^1 + (\mu - \gamma)m^1 - (\bar{\mu} - \bar{\gamma})m^1 - (\beta - \tau)n^1] \\
&- [n^\mu m^j{}_{,\mu} - m^\mu n^j{}_{,\mu} + (\bar{\pi} - \bar{\alpha})n^j + (\mu - \gamma)m^j - (\bar{\mu} - \bar{\gamma})m^j - (\beta - \tau)n^j] r_{H,j}\}(\bar{m}^1 - \bar{m}^j r_{H,j})\}\big|_{r\to r_H}
\end{aligned} ,$$

for $\Psi = F_1$, and

$$C_2 = \frac{1}{g^{01} - g^{0j} r_{H,j}}\bigg|_{r \to r_H} \{\{2 l^{\mu} n^{1}{}_{,\mu} - 2(\pi - \alpha) m^{1} - 2(\bar{\pi} - \bar{\alpha}) \bar{m}^{1} + [(\mu - \gamma) + (\bar{\mu} - \bar{\gamma})] l^{1}$$
$$+ [(\bar{\varepsilon} - \bar{\rho}) + (\varepsilon - \rho)] n^{1} - \left(m^{\mu} \bar{m}^{1}_{,\mu} + \bar{m}^{\mu} m^{1}{}_{,\mu}\right)\}$$
$$-\{2 l^{\mu} n^{j}{}_{,\mu} - 2(\pi - \alpha) m^{j} - 2(\bar{\pi} - \bar{\alpha}) \bar{m}^{j} + [(\mu - \gamma) + (\bar{\mu} - \bar{\gamma})] l^{j}$$
$$+ [(\bar{\varepsilon} - \bar{\rho}) + (\varepsilon - \rho)] n^{j} - \left(m^{\mu} \bar{m}^{j}_{,\mu} + \bar{m}^{\mu} m^{j}{}_{,\mu}\right)\} r_{H,j}\}\Big|_{r \to r_H}$$
$$+ \frac{1}{\left(g^{01} - g^{0j} r_{H,j}\right)\left(l^{j} r_{H,j} - l^{1}\right)}\bigg|_{r \to r_H} \{\{\left[l^{\mu} m^{1}{}_{,\mu} - m^{\mu} l^{1}{}_{,\mu} + (\beta - \tau) l^{1} + (\bar{\varepsilon} - \bar{\rho}) m^{1} - (\varepsilon - \rho) m^{1} - (\bar{\pi} - \bar{\alpha}) l^{1}\right]$$
$$- \left[l^{\mu} m^{j}{}_{,\mu} - m^{\mu} l^{j}{}_{,\mu} + (\beta - \tau) l^{j} + (\bar{\varepsilon} - \bar{\rho}) m^{j} - (\varepsilon - \rho) m^{j} - (\bar{\pi} - \bar{\alpha}) l^{j}\right] r_{H,j}\}\left(\bar{m}^{1} - \bar{m}^{j} r_{H,j}\right)$$
$$+ \{\left[l^{\mu} \bar{m}^{1}_{,\mu} - \bar{m}^{\mu} l^{1}{}_{,\mu} + \left(\bar{\beta} - \bar{\tau}\right) l^{1} + (\varepsilon - \rho) \bar{m}^{1} - (\bar{\varepsilon} - \bar{\rho}) \bar{m}^{1} - (\pi - \alpha) l^{1}\right]$$
$$- \left[l^{\mu} \bar{m}^{j}_{,\mu} - \bar{m}^{\mu} l^{j}{}_{,\mu} + \left(\bar{\beta} - \bar{\tau}\right) l^{j} + (\varepsilon - \rho) \bar{m}^{j} - (\bar{\varepsilon} - \bar{\rho}) \bar{m}^{j} - (\pi - \alpha) l^{j}\right] r_{H,j}\}\left(m^{1} - m^{j} r_{H,j}\right)\}\Big|_{r \to r_H},$$

$$C_1 = \frac{1}{g^{01} - g^{0j} r_{H,j}}\bigg|_{r \to r_H} \{\left\{\frac{[(\mu - \gamma) - (\bar{\mu} - \bar{\gamma})] l^{1}}{2i} + \frac{[(\bar{\varepsilon} - \bar{\rho}) - (\varepsilon - \rho)] n^{1}}{2i} - \frac{\left(m^{\mu} \bar{m}^{1}_{,\mu} - \bar{m}^{\mu} m^{1}{}_{,\mu}\right)}{2i}\right\}$$
$$- \left\{\frac{[(\mu - \gamma) - (\bar{\mu} - \bar{\gamma})] l^{j}}{2i} + \frac{[(\bar{\varepsilon} - \bar{\rho}) - (\varepsilon - \rho)] n^{j}}{2i} - \frac{\left(m^{\mu} \bar{m}^{j}_{,\mu} - \bar{m}^{\mu} m^{j}{}_{,\mu}\right)}{2i}\right\} r_{H,j}\}\bigg|_{r \to r_H}$$
$$+ \frac{1}{\left(g^{01} - g^{0j} r_{H,j}\right)\left(l^{j} r_{H,j} - l^{1}\right)}\bigg|_{r \to r_H} \frac{1}{2i} \{\{\left[l^{\mu} m^{1}{}_{,\mu} - m^{\mu} l^{1}{}_{,\mu} + (\beta - \tau) l^{1} + (\bar{\varepsilon} - \bar{\rho}) m^{1} - (\varepsilon - \rho) m^{1} - (\bar{\pi} - \bar{\alpha}) l^{1}\right]$$
$$- \left[l^{\mu} m^{j}{}_{,\mu} - m^{\mu} l^{j}{}_{,\mu} + (\beta - \tau) l^{j} + (\bar{\varepsilon} - \bar{\rho}) m^{j} - (\varepsilon - \rho) m^{j} - (\bar{\pi} - \bar{\alpha}) l^{j}\right] r_{H,j}\}\left(\bar{m}^{1} - \bar{m}^{j} r_{H,j}\right)$$
$$- \{\left[l^{\mu} \bar{m}^{1}_{,\mu} - \bar{m}^{\mu} l^{1}{}_{,\mu} + \left(\bar{\beta} - \bar{\tau}\right) l^{1} + (\varepsilon - \rho) \bar{m}^{1} - (\bar{\varepsilon} - \bar{\rho}) \bar{m}^{1} - (\pi - \alpha) l^{1}\right]$$
$$- \left[l^{\mu} \bar{m}^{j}_{,\mu} - \bar{m}^{\mu} l^{j}{}_{,\mu} + \left(\bar{\beta} - \bar{\tau}\right) l^{j} + (\varepsilon - \rho) \bar{m}^{j} - (\bar{\varepsilon} - \bar{\rho}) \bar{m}^{j} - (\pi - \alpha) l^{j}\right] r_{H,j}\}\left(m^{1} - m^{j} r_{H,j}\right)\}\Big|_{r \to r_H},$$

for $\Psi = F_2$.

Here the numerator of $I$ is

$$g^{11} - 2 g^{1j} r_{H,j} + g^{jk} r_{H,j} r_{H,k}\Big|_{r \to r_H}.$$

One can see that this is the left side of Eq. (3), so the numerator approaches to zero, and the denominator also approaches to zero. Hence $I$ is an indeterminate form of $\frac{0}{0}$ type. Using L'Hospital's Rule and adjusting $\kappa$ to let $I$ be $1$ [7-9, 12-14, 16-18], then Eq. (14) becomes the standard wave equation in flat spacetime consequently, i.e.

$$I = \frac{\dfrac{\partial g^{11}}{\partial r} - 2\dfrac{\partial g^{1j}}{\partial r} r_{H,j} + \dfrac{\partial g^{jk}}{\partial r} r_{H,j} r_{H,k} + 2\kappa\left[2 g^{11} - 2 g^{0j} r_{H,j}\right]}{2\kappa\left[g^{01} - g^{0j} r_{H,j}\right]}\Bigg|_{r \to r_H} = 1.$$

Thus we can gain

$$\kappa = \frac{\dfrac{\partial g^{11}}{\partial r} + \dfrac{\partial g^{jk}}{\partial r} r_{H,j} r_{H,k} - 2\dfrac{\partial g^{j1}}{\partial r} r_{H,j}}{2\left[g^{01} - 2 g^{11} + \left(2 g^{j1} - g^{j0}\right) r_{H,j}\right]}\Bigg|_{r \to r_H}, \qquad (15)$$

which is the "surface gravity" of event horizon.

By the same reason, $C_{3*}$ is also an indeterminate form of $\frac{0}{0}$ type. According to the similar method we can obtain

$$-2\kappa \lim_{r\to r_H} \frac{g^{11}-2g^{1j}r_{H,j}+g^{jk}r_{H,j}r_{H,k}}{2\kappa(r-r_H)[g^{01}[2\kappa(r-r_H)+1]-g^{0j}r_{H,j}]} = -2\kappa\left[1-2\frac{g^{11}-g^{j1}r_{H,j}}{g^{01}-g^{0j}r_{H,j}}\right]\Bigg|_{r\to r_H} . \tag{16}$$

Then the united form of second order Dirac equations (14) near the event horizon becomes the form as follows

$$\frac{\partial^2\Psi}{\partial {r_*}^2}+2\frac{\partial^2\Psi}{\partial r_*\partial v_*}+B\frac{\partial^2\Psi}{\partial r_*\partial\theta_*}+C\frac{\partial^2\Psi}{\partial r_*\partial\varphi_*}+A\frac{\partial\Psi}{\partial r_*}=0\,, \tag{17}$$

where

$$A=2i(C_0+C_1)+C_2+C_3+C_4\,,$$

$$C_3=-2\kappa\left[1-2\frac{g^{11}-g^{j1}r_{H,j}}{g^{01}-g^{0j}r_{H,j}}\right]\Bigg|_{r\to r_H} .$$

Eq. (17) can be viewed as a usual differential equation because all coefficients in it are regarded as finite real constants. Thus it can be treated by separating variables as [9, 12-14, 16-18]

$$\Psi = R(r_*,v_*)\eta(v_*,\theta_*,\varphi_*)e^{-i\omega v_*+ik_\theta\theta_*+ik_\varphi\varphi_*}\,, \tag{18}$$

where $\eta(v_*,\theta_*,\varphi_*)$ is an arbitrary real function, $\omega$ is the energy of the Dirac particle, $k_\theta, k_\varphi$ are components of generalized momentum of Dirac particles. And we define $k_\theta\equiv P_2\equiv\frac{\partial S}{\partial\theta_*}$, $k_\varphi\equiv P_3\equiv\frac{\partial S}{\partial\varphi_*}$, where $S$ is the Hamiltonian main function of Dirac particles.

Substituting Eq. (18) into Eq. (17), after separating variables one can gain two independent solutions

$$\Psi_{in}=e^{\frac{1}{2}\int[\xi-\lambda(v_*)]dv_*}\eta(v_*,\theta_*,\varphi_*)\exp\left(-i\omega v_*+ik_\theta\theta_*+ik_\varphi\varphi_*\right), \tag{19}$$

$$\Psi_{out}=e^{\frac{1}{2}\int[\xi-\lambda(v_*)]dv_*}\eta(v_*,\theta_*,\varphi_*)\exp\left(2i\omega-A-ik_\theta B-ik_\varphi C-\xi\right)r_*\cdot\exp\left(-i\omega v_*+ik_\theta\theta_*+ik_\varphi\varphi_*\right), \tag{20}$$

where $\xi$ is a constant in variable separation, and $\lambda(v_*)$ is a function of the advanced time $v_*$ in variable separation, where $\lambda(v_*)=2\frac{\frac{\partial\eta}{\partial v_*}}{\eta}+B\frac{\frac{\partial\eta}{\partial\theta_*}}{\eta}+C\frac{\frac{\partial\eta}{\partial\varphi_*}}{\eta}$ and $\xi=\frac{2T'+\lambda(v_*)T}{T}$ (we let $R(r_*,v_*)=P(r_*)T(v_*)$ in variable separation).

The radial components of the above two independent solutions are

$$\psi_{in}=e^{\frac{1}{2}\int[\xi-\lambda(v_*)]dv_*}\exp\left(-i\omega v_*\right), \tag{21}$$

$$\begin{aligned}\psi_{out}&=e^{\frac{1}{2}\int[\xi-\lambda(v_*)]dv_*}\exp\left(2i\omega-A-ik_\theta B-ik_\varphi C-\xi\right)r_*\cdot\exp\left(-i\omega v_*\right)\\&=e^{\frac{1}{2}\int[\xi-\lambda(v_*)]dv_*}e^{-i\omega v_*}e^{2i\left(\omega-k_\theta\frac{B}{2}-k_\varphi\frac{C}{2}-\frac{\mathrm{Im}(A)}{2}\right)r_*}e^{[-\xi-\mathrm{Re}(A)]r_*}\end{aligned}\,. \tag{22}$$

Because $\psi_{out}$ is not analytic at the event horizon, it needs extending by analytic continuation into the event horizon through the lower-half complex $r$-plane as [5, 6, 31]

$$(r-r_H)\to|r-r_H|e^{-i\pi}=(r_H-r)e^{-i\pi}\,. \tag{23}$$

Hence

$$\widetilde{\psi}_{out} = \left[ e^{\frac{1}{2}\int[\xi-\lambda(v_*)]dv_*} e^{-i\omega v_*} e^{2i\left(\omega-k_\theta\frac{B}{2}-k_\varphi\frac{C}{2}-\frac{\mathrm{Im}(A)}{2}\right)r_*} e^{[-\xi-\mathrm{Re}(A)]r_*} \right] \cdot e^{\frac{\pi}{\kappa}\left(\omega-k_\theta\frac{B}{2}-k_\varphi\frac{C}{2}-\frac{\mathrm{Im}(A)}{2}\right)} \cdot e^{i\pi\frac{\mathrm{Re}(A)+\xi}{2\kappa}} . \quad (24)$$

From Eq. (22) and Eq. (24) we can obtain the relative scattering probability at the event horizon

$$\left|\frac{\psi_{out}}{\widetilde{\psi}_{out}}\right|^2 = e^{-\frac{2\pi}{\kappa}(\omega-\omega_0)}, \quad (25)$$

where

$$\omega_0 = k_\theta\frac{B}{2} + k_\varphi\frac{C}{2} + \frac{\mathrm{Im}(A)}{2}. \quad (26)$$

Following Damour, Ruffini [5] and Sannan [6], we can get the thermal radiation spectrum of Dirac particles (or scalar particles) from general nonstationary black holes

$$N_\omega = \frac{1}{e^{\frac{\omega-\omega_0}{k_B T}} \pm 1}, \quad (27)$$

and the Hawking temperature

$$T = \frac{\kappa}{2\pi k_B}, \quad (28)$$

where $k_B$ is Boltzmann constant, "$\pm$" correspond to fermion and boson, respectively. The temperature $T$ is

$$T = \frac{\kappa}{2\pi k_B} = \left.\frac{\frac{\partial g^{11}}{\partial r} + \frac{\partial g^{jk}}{\partial r} r_{H,j} r_{H,k} - 2\frac{\partial g^{j1}}{\partial r} r_{H,j}}{4\pi k_B\left[g^{01} - 2g^{11} + \left(2g^{j1} - g^{j0}\right) r_{H,j}\right]}\right|_{r\to r_H}. \quad (29)$$

And the chemical potential is

$$\omega_0 = k_\theta \left.\frac{g^{12} - g^{j2} r_{H,j}}{g^{01} - g^{0j} r_{H,j}}\right|_{r\to r_H} + k_\varphi \left.\frac{g^{13} - g^{j3} r_{H,j}}{g^{01} - g^{0j} r_{H,j}}\right|_{r\to r_H} + e\left.\frac{A_\mu g^{\mu 1} - A_\mu g^{\mu j} r_{H,j}}{g^{01} - g^{0j} r_{H,j}}\right|_{r\to r_H} + C_1 . \quad (30)$$

We know from Eq. (29) that $T$ depends on the time and different angles, so it is a distribution of temperatures. It is also interesting to find that the coefficient $C_1$ appears in the chemical potential, which may represent a particular energy term for Dirac particles. The physical meaning of $C_1$ will be discussed in section 4.

## 3. Research on non-thermal radiation of general nonstationary black holes

Now we use the methods and conclusions of references [22-28] to investigate the non-thermal radiation.

Considering the Hamilton-Jacobi equation of moving particles

$$g^{\mu\nu}\left(\frac{\partial S}{\partial x^\mu} + eA_\mu\right)\left(\frac{\partial S}{\partial x^\nu} + eA_\nu\right) + {\mu_0}^2 = 0. \quad (31)$$

Applying the generalized tortoise coordinate transformation to Eq. (31), via some simplifications, we let the solution of $\frac{\partial S}{\partial r_*}$ be real, and define $\frac{\partial S}{\partial v_*} \equiv -\omega, \frac{\partial S}{\partial \theta_*} \equiv P_2 \equiv k_\theta$, $\frac{\partial S}{\partial \varphi_*} \equiv P_3 \equiv k_\varphi$. Finally we can obtain the energy level distribution of the particles

$$\omega \geq \omega^{+}, \ \ \omega \leq \omega^{-}. \tag{32}$$

When $r \to r_H$, we have

$$\omega^{\pm}\Big|_{r\to r_H} = \omega_0 = k_\theta \frac{g^{12} - g^{j2} r_{H,j}}{g^{01} - g^{0j} r_{H,j}}\Bigg|_{r\to r_H} + k_\varphi \frac{g^{13} - g^{j3} r_{H,j}}{g^{01} - g^{0j} r_{H,j}}\Bigg|_{r\to r_H} + e\frac{A_\mu g^{\mu 1} - A_\mu g^{\mu j} r_{H,j}}{g^{01} - g^{0j} r_{H,j}}\Bigg|_{r\to r_H}. \tag{33}$$

$\omega_0$ is the maximum value of negative energy state after energy level overlapping near the event horizon. Therefore, the incident negative energy particles satisfying $\mu < \omega \leq \omega_0$ will become emerging positive energy particles via quantum tunneling effect. This is the non-thermal radiative process that has no relation to the temperature.

## 4. Discussion

The Hamilton-Jacobi equation describes the general behavior of moving particles in a non-statistical way, so it is suitable for investigating the non-thermal radiation.

The thermal radiation spectrum (27) demonstrates that the total interaction energy of Dirac particles in a general nonstationary black hole is

$$\omega_0 = k_\theta \frac{g^{12} - g^{j2} r_{H,j}}{g^{01} - g^{0j} r_{H,j}}\Bigg|_{r\to r_H} + k_\varphi \frac{g^{13} - g^{j3} r_{H,j}}{g^{01} - g^{0j} r_{H,j}}\Bigg|_{r\to r_H} + e\frac{A_\mu g^{\mu 1} - A_\mu g^{\mu j} r_{H,j}}{g^{01} - g^{0j} r_{H,j}}\Bigg|_{r\to r_H} + C_1 \ , \tag{34}$$

where

$$\begin{aligned}
C_1 = {} & \frac{1}{g^{01} - g^{0j} r_{H,j}}\Bigg|_{r\to r_H} \{\left\{\frac{[(\bar\mu - \bar\gamma) - (\mu - \gamma)]l^1}{2i} + \frac{[(\varepsilon - \rho) - (\bar\varepsilon - \bar\rho)]n^1}{2i} - \frac{\left(\bar m^\mu m^1{}_{,\mu} - m^\mu \bar m^1_{,\mu}\right)}{2i}\right\} \\
& - \left\{\frac{[(\bar\mu - \bar\gamma) - (\mu - \gamma)]l^j}{2i} + \frac{[(\varepsilon - \rho) - (\bar\varepsilon - \bar\rho)]n^j}{2i} - \frac{\left(\bar m^\mu m^j{}_{,\mu} - m^\mu \bar m^j_{,\mu}\right)}{2i}\right\} r_{H,j}\}\Bigg|_{r\to r_H} \\
& + \frac{1}{\left(g^{01} - g^{0j} r_{H,j}\right)\left(n^j r_{H,j} - n^1\right)}\Bigg|_{r\to r_H} \frac{1}{2i}\{\{\left[n^\mu \bar m^1_{,\mu} - \bar m^\mu n^1{}_{,\mu} + (\pi - \alpha) n^1 + (\bar\mu - \bar\gamma)\bar m^1 - (\mu - \gamma)\bar m^1 - \left(\bar\beta - \bar\tau\right) n^1\right] \\
& - \left[n^\mu \bar m^j_{,\mu} - \bar m^\mu n^j{}_{,\mu} + (\pi - \alpha) n^j + (\bar\mu - \bar\gamma)\bar m^j - (\mu - \gamma)\bar m^j - \left(\bar\beta - \bar\tau\right) n^j\right] r_{H,j}\}\left(m^1 - m^j r_{H,j}\right) \\
& - \{\left[n^\mu m^1{}_{,\mu} - m^\mu n^1{}_{,\mu} + (\bar\pi - \bar\alpha) n^1 + (\mu - \gamma) m^1 - (\bar\mu - \bar\gamma) m^1 - (\beta - \tau) n^1\right] \\
& - \left[n^\mu m^j{}_{,\mu} - m^\mu n^j{}_{,\mu} + (\bar\pi - \bar\alpha) n^j + (\mu - \gamma) m^j - (\bar\mu - \bar\gamma) m^j - (\beta - \tau) n^j\right] r_{H,j}\}\left(\bar m^1 - \bar m^j r_{H,j}\right)\}\Big|_{r\to r_H}
\end{aligned} \tag{35}$$

for $\Psi = F_1$, and

$$\begin{aligned}
C_1 = {} & \frac{1}{g^{01} - g^{0j} r_{H,j}}\Bigg|_{r\to r_H} \{\left\{\frac{[(\mu - \gamma) - (\bar\mu - \bar\gamma)]l^1}{2i} + \frac{[(\bar\varepsilon - \bar\rho) - (\varepsilon - \rho)]n^1}{2i} - \frac{\left(m^\mu \bar m^1_{,\mu} - \bar m^\mu m^1{}_{,\mu}\right)}{2i}\right\} \\
& - \left\{\frac{[(\mu - \gamma) - (\bar\mu - \bar\gamma)]l^j}{2i} + \frac{[(\bar\varepsilon - \bar\rho) - (\varepsilon - \rho)]n^j}{2i} - \frac{\left(m^\mu \bar m^j_{,\mu} - \bar m^\mu m^j{}_{,\mu}\right)}{2i}\right\} r_{H,j}\}\Bigg|_{r\to r_H} \\
& + \frac{1}{\left(g^{01} - g^{0j} r_{H,j}\right)\left(l^j r_{H,j} - l^1\right)}\Bigg|_{r\to r_H} \frac{1}{2i}\{\{\left[l^\mu m^1{}_{,\mu} - m^\mu l^1{}_{,\mu} + (\beta - \tau) l^1 + (\bar\varepsilon - \bar\rho) m^1 - (\varepsilon - \rho) m^1 - (\bar\pi - \bar\alpha) l^1\right] \\
& - \left[l^\mu m^j{}_{,\mu} - m^\mu l^j{}_{,\mu} + (\beta - \tau) l^j + (\bar\varepsilon - \bar\rho) m^j - (\varepsilon - \rho) m^j - (\bar\pi - \bar\alpha) l^j\right] r_{H,j}\}\left(\bar m^1 - \bar m^j r_{H,j}\right) \\
& - \{\left[l^\mu \bar m^1_{,\mu} - \bar m^\mu l^1{}_{,\mu} + \left(\bar\beta - \bar\tau\right) l^1 + (\varepsilon - \rho)\bar m^1 - (\bar\varepsilon - \bar\rho)\bar m^1 - (\pi - \alpha) l^1\right] \\
& - \left[l^\mu \bar m^j_{,\mu} - \bar m^\mu l^j{}_{,\mu} + \left(\bar\beta - \bar\tau\right) l^j + (\varepsilon - \rho)\bar m^j - (\bar\varepsilon - \bar\rho)\bar m^j - (\pi - \alpha) l^j\right] r_{H,j}\}\left(m^1 - m^j r_{H,j}\right)\}\Big|_{r\to r_H}
\end{aligned} \tag{36}$$

for $\Psi = F_2$.

This chemical potential $\omega_0$ is composed of three parts. The sum of the first two terms $k_\theta \left. \frac{g^{12} - g^{j2} r_{H,j}}{g^{01} - g^{0j} r_{H,j}} \right|_{r \to r_H} + k_\varphi \left. \frac{g^{13} - g^{j3} r_{H,j}}{g^{01} - g^{0j} r_{H,j}} \right|_{r \to r_H}$ is the rotational energy arising from the coupling between different components of generalized momentum of particles and different rotations of the black hole. The second part is the electro-magnetic interaction energy $e \left. \frac{A_\mu g^{\mu 1} - A_\mu g^{\mu j} r_{H,j}}{g^{01} - g^{0j} r_{H,j}} \right|_{r \to r_H}$. The third part $C_1$ ( which characterizes a new extra coupling effect arising from the interaction between the intrinsic spin of particles and the generalized momentum of the evaporating black hole and vanishes in the case of scalar particles (we will see it later) and has no classical correspondence ) gives a general expression of the new extra spin-rotation coupling and the spin-acceleration coupling effect. When considering concrete black holes like those of Ref. [16-18], $C_1$ will degenerate and represent the concrete extra coupling effects in Ref. [16-18].

Now we consider the asymptotic behavior of minimally electromagnetic coupling Klein-Gordon equation near the event horizon. The explicit form of wave equation describing the dynamic behavior of scalar particles with mass $\mu_0$ and charge $e$

$$\frac{1}{\sqrt{-g}} \left[ \left( \frac{\partial}{\partial x^\mu} + ieA_\mu \right) \sqrt{-g} g^{\mu\nu} \left( \frac{\partial}{\partial x^\nu} + ieA_\nu \right) \right] \phi(x) - {\mu_0}^2 \phi(x) = 0 \,. \tag{37}$$

Applying the same method in section 2 to Eq. (37), throughout long calculations, one can get a same form of Eq. (17) near the event horizon, except that the coefficient $A = 2iC_0 + C_3 + {C_4}'$, where

$${C_4}' = \left. \frac{\left( {g^{\mu 1}}_{,\mu} + \frac{1}{\sqrt{-g}} g^{\mu 1} \frac{\partial \sqrt{-g}}{\partial x^\mu} \right) - \left( {g^{\mu j}}_{,\mu} + \frac{1}{\sqrt{-g}} g^{\mu j} \frac{\partial \sqrt{-g}}{\partial x^\mu} \right) r_{H,j} - g^{jk} r_{H,jk}}{g^{01} - g^{0j} r_{H,j}} \right|_{r \to r_H} ,$$

and the chemical potential

$$\begin{aligned} \omega_0 &= k_\theta \frac{B}{2} + k_\varphi \frac{C}{2} + \frac{\operatorname{Im}(A)}{2} \\ &= k_\theta \left. \frac{g^{12} - g^{j2} r_{H,j}}{g^{01} - g^{0j} r_{H,j}} \right|_{r \to r_H} + k_\varphi \left. \frac{g^{13} - g^{j3} r_{H,j}}{g^{01} - g^{0j} r_{H,j}} \right|_{r \to r_H} + e \left. \frac{A_\mu g^{\mu 1} - A_\mu g^{\mu j} r_{H,j}}{g^{01} - g^{0j} r_{H,j}} \right|_{r \to r_H} . \end{aligned} \tag{38}$$

One can see that $C_1 = 0$ in this case. Furthermore, it is interesting to find that Eq. (38) and Eq. (33) are the same, which means that the chemical potential in thermal radiation spectrum is equal to the highest energy of negative energy state of scalar particles in non-thermal radiation for general nonstationary black holes. This is an important conclusion that reveals the relationship between the two kinds of radiative processes of black holes, which is ignored before. Eq. (33) is derived from section 3 by using the Hamilton-Jacobi equation which applies to both Fermion and Boson. So Eq. (33) is also for scalar particles. This is why we can compare these two equations.

In fact, "temperature" and "chemical potential" in this letter are analogical concepts of ordinary temperature and chemical potential. Mathematically speaking, in our letter, temperature and chemical potential are functions that describe certain concepts of statistical mechanics. When concerning equilibrium or near-equilibrium conditions (e.g., slowly varying black holes), they will be degenerated to the ordinary concepts of temperature and chemical potential, respectively. Most of the nonstationary black holes are slowly varying, otherwise they can not exist due to rapid evaporation. So, for most of the nonstationary black holes, the concepts of temperature and chemical potential take effect as their ordinary meanings (for example, the chemical potential at zero temperature is equivalent to the Fermi energy of a fermions system, which is a kind of boundary for energy levels), and the thermal radiation and non-thermal radiation have the forenamed relationship that the chemical potential (as its ordinary meaning) in thermal radiation spectrum is equal to the highest energy of negative energy state of scalar particles in non-thermal radiation.

On the other hand, because a lot of general physical processes should satisfy quantitative causal relation with no-loss-no-gain character [32-34], e.g., Ref. [35] uses the no-loss-no-gain homeomorphic map transformation satisfying the quantitative causal relation to gain exact strain tensor formulas in Weitzenböck manifold. In fact, some changes ( cause ) of some quantities in (7) must result in the relative some changes ( result ) of the other quantities in (7) so that (7)'s right side keep no-loss-no-gain, i.e., zero, namely, (7) also satisfies the quantitative causal relation. And (2), (8-11), (31), (37) also satisfy the quantitative causal relation in the same way. Hence the researches in this letter are consistent. Also, the researches of this letter provide an alternative and convenient way to obtain the Hawking temperature.

## 5. Summary and Conclusion

This letter carefully investigates quantum radiation of Dirac particles in general nonstationary black holes, generally shows that the temperature and shape of event horizon of black holes depend on both the time and different angles in general condition, and further obtains a general expression of the new extra coupling effect arising from the interaction between the intrinsic spin of Dirac particles and the generalized momentum of the general nonstationary black hole. Finally, this letter shows that the new extra coupling effect is absent in the thermal radiation spectrum in this case, and generally reveals a relationship that is ignored before between thermal radiation and non-thermal radiation of black holes in detail, which is that the chemical potential in thermal radiation spectrum is equal to the highest energy of the negative energy state in non-thermal radiation for general nonstationary black holes.

**Acknowledgment**

The authors are grateful for Prof. R. G. Cai and Prof. Z. Zhao for their useful discussion and comment. The work is supported by National Natural Science Foundation of China (No. 10875009 ) and Beijing Natural Science Foundation (No.1072005).